\documentclass{article}
\usepackage[utf8]{inputenc}
\usepackage[numbers]{natbib}
\usepackage{graphicx}
\usepackage[letterpaper]{geometry}  
\usepackage{subcaption}  
\usepackage{hyperref}
\hypersetup{colorlinks=true,urlcolor=blue,citecolor=brown}  
\usepackage{authblk}

\usepackage{xcolor}  

\title{Propensity score models are better when post-calibrated}
\author[1,2]{Rom Gutman}
\author[1]{Ehud Karavani\thanks{Correspondence to ehudk@ibm.com.}}
\author[1]{Yishai Shimoni}
\affil[1]{IBM Research, Israel}
\affil[2]{Technion - Israel Institute of Technology, Haifa, Israel}

\date{}

\begin{document}
\maketitle

\begin{abstract}
    Theoretical guarantees for causal inference using propensity scores are partly based on the scores behaving like conditional probabilities.
    However, scores between zero and one, especially when outputted by flexible statistical estimators, do not necessarily behave like probabilities.
    We perform a simulation study to assess the error in estimating the average treatment effect before and after applying a simple and well-established post-processing method to calibrate the propensity scores. 
    We find that post-calibration reduces the error in effect estimation for expressive uncalibrated statistical estimators,
    and that this improvement is not mediated by better balancing.
    The larger the initial lack of calibration, the larger the improvement in effect estimation,
    with the effect on already-calibrated estimators being very small.
    Given the improvement in effect estimation and that post-calibration is computationally cheap,
    we recommend it will be adopted when modelling propensity scores with expressive models.
\end{abstract}

Keywords: causal inference, propensity score, calibration, model validation, average treatment effect.

\section{Introduction}


The propensity score is defined as the conditional probability of being assigned to a treatment (exposure) given one's observed confounding variables. 
It is very commonly used in methods for estimating causal effects from observational data, such as inverse probability weighting \cite{robins2000marginal}, propensity matching \cite{rosenbaum1983propensity, rosenbaum1985constructing}, propensity stratification \cite{rosenbaum1984reducing}, as well as many doubly-robust methods \cite{kang2007demystifying,schuler2017targeted, bang2005doubly, glynn2010introduction}

\citet{rosenbaum1983propensity} set up theoretical guaranties ensuring that adjusting for the propensity score, instead of the covariates themselves, is sufficient in order to achieve the conditional exchangeability needed to estimate a causal effect.
However, while these theoretical guarantees require the true conditional probabilities,
when applied in practice, not every model that inputs data and outputs a number between zero and one, correctly estimates true probabilities.
The scores might not reliably represent true probabilities.


A prediction model that accurately outputs probabilities is referred to as calibrated  
(note this is unrelated to a previous notion of "propensity score calibration" from \cite{sturmer2007performance}).
Calibration can be empirically evaluated with calibration curve (reliability curves), comparing the predicted scores with their corresponding rate of labels \citep{zadrozny2002transforming}. 
Formally, for a model $f$ that regresses binary outcome $Y$ on covariates $X$, calibration is defined as $E[Y|f(x)]=f(x)$.
To illustrate with an example, under this notion of empirical probability, taking all the observations for which the model predicted a score of 0.8 should result in about 80\% of them having a positive label.
Since an entire curve is not always actionable, there are also multiple metrics that try to capture this notion with a single numerical value \citep{huang2020tutorial, harrell2001regression, austin2019ici}.

While papers using propensity-based methods might check for overlapping propensity distributions or covariate balancing \cite{tazare2022transparency, granger2020review, shimoni2019evaluation}, 
it is somewhat uncommon for them to evaluate their propensity scores for calibration.  
Fortunately, since a majority of classical statistics literature tends to estimate propensity scores using logistic regression models, they might have inadvertently overcome the need.
Logistic regression models are fitted by optimizing the log-loss objective function (binary cross-entropy, negative Shanon entropy) \citep{jordan1995logistic, eslii},
which is, by itself, a metric for calibration \citep{gneiting2007strictly}.
Therefore, well-specified estimators optimizing for it usually result in well-calibrated models.

However, not all statistical estimators are inherently calibrated.
Higher-complexity models, such as tree-based or neural-network-based models, 
might not get calibration for free like logistic regression does.
Even regularizing logistic regression (LASSO, ridge regression, or elastic-net models) might harm calibration, as the penalty added moves the objective function away from the "pure" and calibrating log-loss \citep{vancalster2020shrinkage, vsinkovec2021tune}.
As these models become more popular for propensity estimation \citep{westreich2010propensity},
it should be of interest to know whether calibration of propensity models is important for effect estimation.

Furthermore, it is of interest whether we can break the trade-off between model-expressiveness and calibration by post-calibrating estimators.
Therefore, post-calibration may hold a promise for using complex high-dimensional data for propensity score estimation when performing causal inference from observational data. 

In this paper we will use simulations to quantify the downstream effect of poorly-calibrated conditional probabilities on estimation of causal effects.
We hypothesize that well-calibrated propensities are indeed imperative to properly estimate causal effects, 
and that in cases where calibration is poor, 
effect estimation will be improved by post-calibrating the propensity models.

\section{Methods} \label{sec:methods}
To assess the importance of calibration we use simulations, so we have access to individual-level propensity scores and counterfactual outcomes. 
Below we describe the data generating processes used, the method for effect estimation, and the measurements obtained from the various estimations.

\subsection{Causal inference framework}
We denote the binary treatment assignment for each individual $i$ as $A_i$, 
the covariates (ideally confounders) as $X_i$, 
and the true propensity to be treated $\pi_i = \Pr[A_i=1|X=x_i]$.
Using Rubin's potential outcomes framework \citep{rubin1974estimating}, we notate $Y_i^{A=a}$ as the hypothetical outcome that would have been observed had individual $i$ received treatment $a$, 
and, assuming consistency, the observed outcome is the one corresponding to the treatment actually assigned $Y_i=Y_i^{A=a_i}$.
Lastly, we define the average treatment effect (ATE) as $E[Y^1 - Y^0]$.

\subsection{Estimation}
To estimate the causal effect from the observed data we first estimate the propensity score, denoted as $\hat{\pi}$, by regressing the treatment assignment on the covariates. 
We fit various estimators, common in the literature: logistic regression, regularized logistic regression - LASSO and Ridge, random forest, and gradient boosting trees (additive trees).

Since these models require hyperparameters, we perform hyperparameter search using cross validation and select the configuration maximizing Brier score \citep{brier1950score}, which is a calibration-inducing metric. 
Trees are fine-tuned for their depth and number of trees in the ensemble, and regularized logistic regression are fine-tuned for the strength of regularization. 
To avoid overfit, the prediction of the propensity scores is done on unseen data points using cross validation.
Hence, overall a nested cross-validation approach is taken with inner cross validation used for parameter estimation and the outer one for estimation.

Once propensity scores are obtained, we plug them into an inverse probability weighting (IPW) estimator \citep{horvitz1952generalization} to estimate the average treatment effect.

\subsubsection{Post-calibration}
Optionally, we can post-process the predicted propensity scores by calibrating them before using them with IPW.
There are multiple methods to perform this post-calibration that should result in scores functioning more like probabilities. 
In this study we focus on Platt's scaling \citep{platt1999probabilistic} as it is most appropriate for the data generating process used.
The method takes advantage of the well-calibrated properties of log-loss by fitting a logistic regression over the scores outputted from an estimator against the treatment assignment labels.

\subsection{Measurements}
For each set of propensity scores and post-calibrated propensity scores we take four main measurements.

First, we measure the calibration error.
This is done graphically with calibration curves and numerically with Integrated Calibrated Index (ICI) \citep{austin2019ici}.
Calibration curves present the notion of empirical calibration.
It bins the predicted scores, and in each such bin it counts the number of positive and negative labels.
A well-calibrated model will have the same rate of observed labels as the average score of the bin, thus resulting in a diagonal line along the x=y curve.
ICI is a way to extract a numeric value from the notion of the diagonal calibration curve.
It fits a LOESS regression between the binary classes and the predicted scores, 
calculates the difference between the resulting predicted line and the optimal x=y diagonal, and takes the mean of those absolute differences.

Second, we measure the effect estimation error.
The propensity scores are transformed into inverse propensity weights and the effect is estimated.
We then take the absolute difference between the ground truth effect from the simulation and the estimated one from the IPW.

Third, we measure covariate balance between the treatment groups.
We calculate the absolute standardized mean difference, after inverse propensity weighting, for each covariate and then take the maximum value over all covariates. 

\subsection{data}
\subsubsection{Simulation data}
To estimate the effect of miscalibration on effect estimation we first use an estimation-free propensity scores simulation.
We apply the following simple data generating process:

$$
\pi = expit(\gamma (-0.1 X_1 + 0.05 X2 + 0.2 X_3 - 0.05 X_4 + \epsilon))
$$
$$
A = Bernoulli(\pi)
$$
$$
Y^a = 5 A + 1.2 X_1 + 3.6 X_2 + 1.2 X_3 + 1.2 X_4 + \epsilon'
$$

Where $X \sim \mathcal{N}(0, 3^2)$ and $\epsilon, \epsilon' \sim \mathcal{N}(0, 0.5^2)$

We set $\gamma=1$ while generating the data.
We imitate uncalibrated models by purposefully decalibrating the true propensity.
We do so by scaling $\pi$ into $\hat{\pi}$ for different $\gamma$ values in the logarithmic range of $(0.25,3)$.
For each $\gamma$ value we generate 10 repetitions, each of size 10,000.

We also used data from the $\gamma=1$ setting when fitting the statistical estimators, whose results appear in the \nameref{sec:supp_mat}.

\subsubsection{ACIC 2016 data}
To estimate the impact of post-calibration on different statistical estimators in a more realistic scenario, we also use a more complex data generation process used in the 2016 Atlantic Causal Inference Conference Data Challenge.
The data is semi-synthetic based on real covariates from the Collaborative Perinatal Project longitudinal study, and is used to simulate treatment assignments and potential outcomes.
There are multiple generating processes and multiple realizations for each process,
but each dataset has the same 4802 observations and of 58 covariates.
Complete details on the data generating process can be found in \citep{dorie2019automated}.
For this study we selected five instances of a single data generating process (numbered 42) with a relatively simple response structure and decent covariate overlap. 

As above, we obtain the propensity scores for each model, use them to estimate the average causal effect with IPW, and take multiple measurements.
We then post-calibrate the propensity scores and repeat. 

All code and results are available at Github: \url{https://github.com/RomGutman/propensity_calibration}

\section{Results}
\subsection{Estimation-free propensity scores}
In the first experiment we check the impact of calibration on the downstream effect estimation by
simulating synthetic treatment assignment and potential outcomes. 
We take the true propensities-to-treat and gradually de-calibrate them through scaling. 
We first use the de-calibrated propensities to estimate the effect with inverse probably weighting (IPW). 
Then we re-calibrate the de-calibrated propensities using Platt's correction \citep{platt1999probabilistic} and estimate the effect again.
This allows us to examine the downstream effect of de-calibration and post-calibration in a tightly controlled setting, with dose-response-like intervention while holding many other "real-world" degrees of freedom constant.

\begin{figure}[h]
    \captionsetup{skip=2pt}  
    \centering
    \begin{subfigure}{0.49\linewidth}
        \centering
        \includegraphics[width=\linewidth]{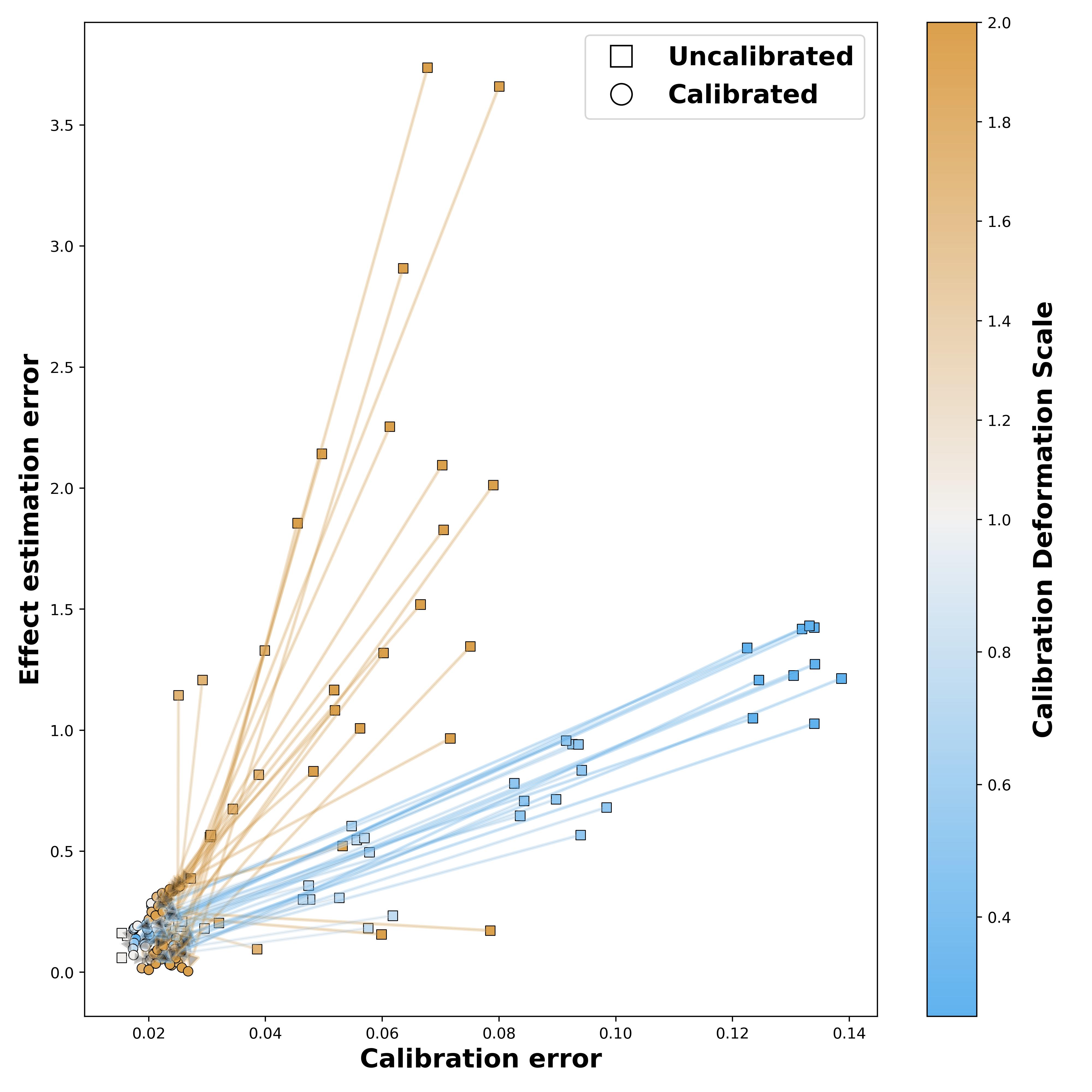}
        \caption{}
        \label{fig:sim_cal_ate}
    \end{subfigure}
    \begin{subfigure}{0.49\linewidth}
        \centering
        \includegraphics[width=\linewidth]{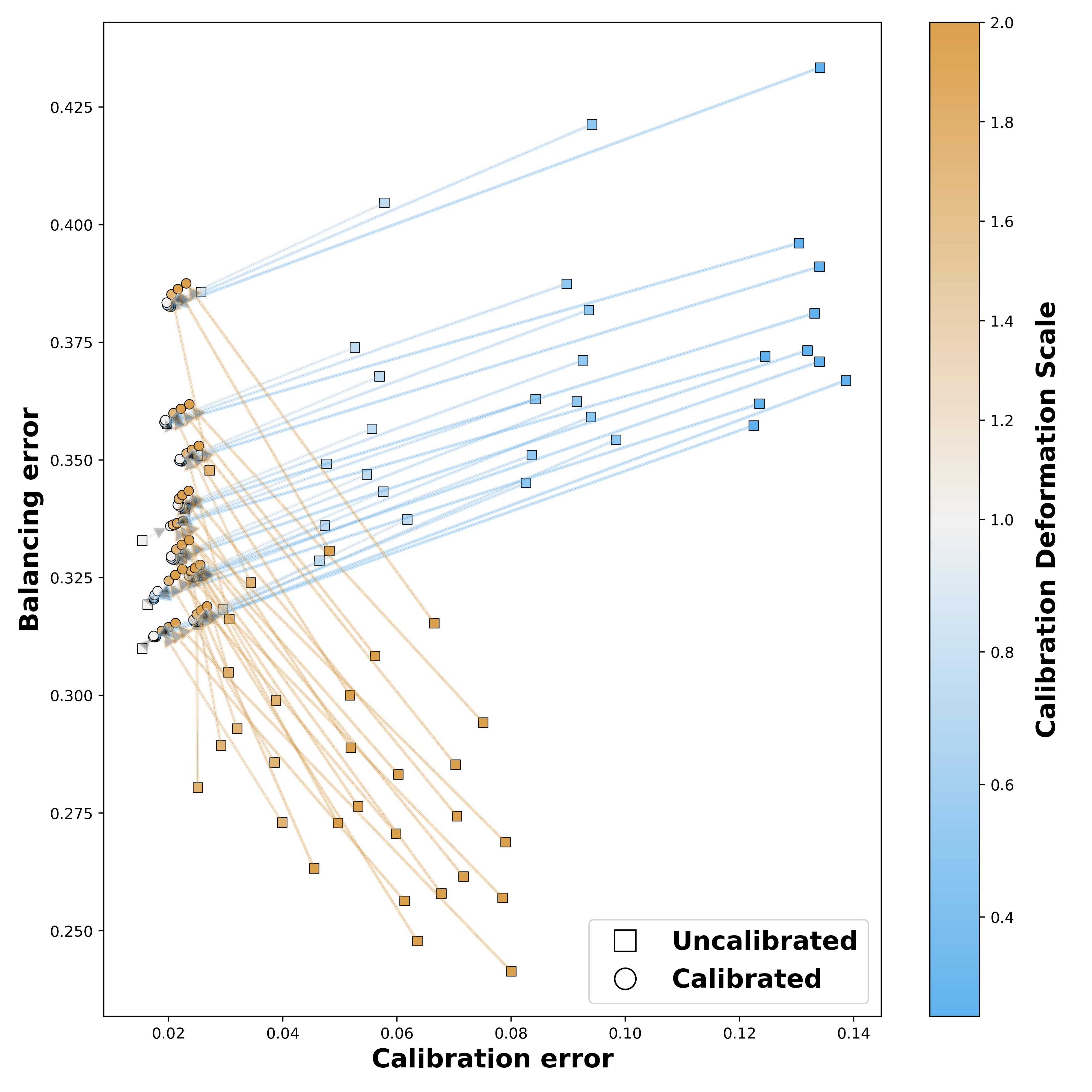}
        \caption{}
        \label{fig:sim_cal_bal}
    \end{subfigure}
    \caption{Post-calibration improves effect estimation, independent of its effect on balancing.
    Using synthetic data with known effect sizes and propensities-to-treat,
    propensity scores were deformed to de-calibrate them and then were re-calibrated. 
    Magnitude of the deformation is coded in color (the further from 1 the larger the deformation). 
    Left panel plots the errors in effect estimation and calibration before and after post-calibration with arrows connecting corresponding pairs. Prior to post-calibration (rectangles), the bigger the propensity deformation (more saturated color) the larger the calibration and effect estimation errors are. After post-calibration (circles) both the calibration error (mean Integrated Calibrated Index) and the average treatment effect estimation error (absolute difference between true and estimated effect) decreases. 
    The right panel shows the effect of calibration on covariate balancing (maximum value of absolute standardized mean difference among covariates), showing that post-calibration does not improve balancing consistently. Therefore suggesting that the improvement in effect estimation is not mediated by improvement in balancing.
    }
    \label{fig:sim_cal_deform}
\end{figure}

Figure \ref{fig:sim_cal_deform} (a) shows the error in effect estimation increases with the increasing magnitude of de-calibration of the true propensity to treat.
It further shows that re-calibrating the deformed propensities improves the effect estimation.

We can quantify the improvement by taking average slopes between de-calibrated and re-calibrated points in the plane stretched by the calibration error and the effect estimation error. 
Table \ref{table:sim_deform} shows two things over a 1000 repetitions per deformation scale. 
First, the magnitude of the slope increases the stronger the de-calibration is.
Second, the sign of the slop is consistently negative, meaning post-calibration consistently improves in both reducing the calibration error and the effect estimation error.
Additionally, while calibration reduced effect estimation error, Figure \ref{fig:sim_cal_deform} (b) shows that it did not consistently improve balancing between the treatment and control groups.

\begin{figure}
    \centering
    \includegraphics[width=\textwidth]{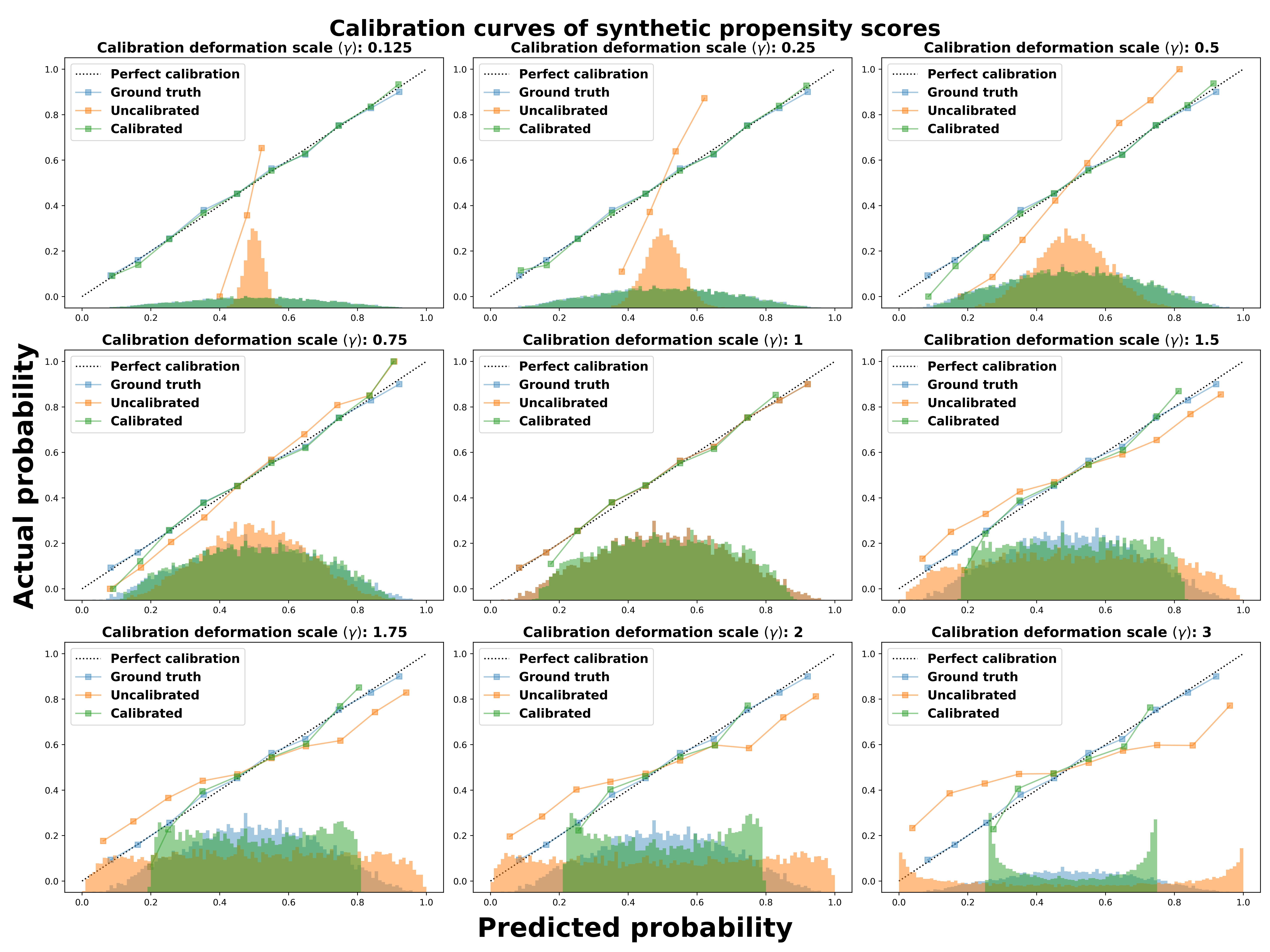}
    \caption{The larger the de-calibration the larger the deviation from optimal x=y diagonal of the calibration curves.
    The figure shows the propensity scores distribution and calibration curves for different deformation scales.
    The larger the deformation is (the further the scale is from 1), the larger the deviation from the diagonal (orange line). 
    Post calibrating the scores results in a a calibration curve (green) closer to the curve defined by the true (before deformation) curve (blue) which is very close to the optimal diagonal curve (dashed line).
    }
    \label{fig:sim_cal_curve}
\end{figure}

We can further look one step deeper into the actual calibration curves of one instance from each deformation scale. 
Figure \ref{fig:sim_cal_curve} shows the effect of the deformation on the distribution of propensities and calibration slopes. 
The deformation magnitude controls the kurtosis of the propensity distribution, with small values leading to high kurtosis and large values lead to uniformly-looking low kurtosis propensity distributions.
It further shows that as the deformation magnitude increases, i.e. moves further away from 1, calibration indeed deteriorates (orange line), moving further from the x=y diagonal (dashed line).
However, correcting it with post-calibration (green line) improves calibration, getting it closer to the diagonal and the curve determined by the true propensity scores (blue line) and the optimal diagonal (dashed line).

\subsection{Model-estimated propensity scores}

In the second experiment, we examine the effect of post-calibrating different propensity estimators on the effect estimation, using more complex semi-synthetic data.
We fit different propensity models using random forests, gradient boosting trees (aka additive trees), logistic regressions and regularized logistic regression (LASSO and ridge), and use IPW to obtain the difference between the estimated potential outcomes.
We measure the calibration error and the effect estimation error, then post-calibrate the propensity scores and measure again.

\begin{figure}[h]
    \centering
    \captionsetup{skip=2pt}  
    \begin{subfigure}[t]{0.49\linewidth}
        \centering
        \includegraphics[width=\linewidth]{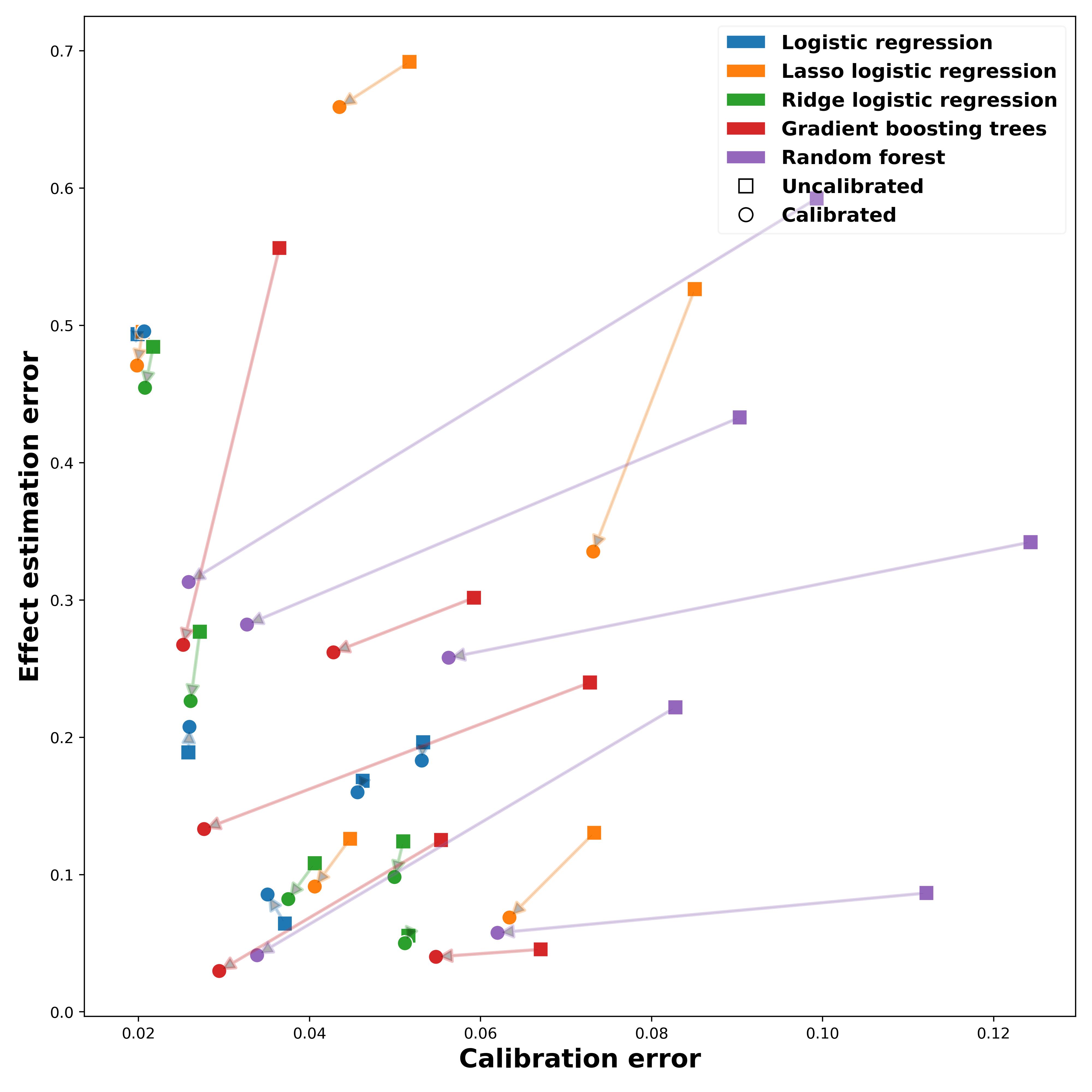}
        \caption{}
        \label{fig:acic_cal_ate}
    \end{subfigure}
    \begin{subfigure}[t]{0.49\linewidth}
        \centering
        \includegraphics[width=\linewidth]{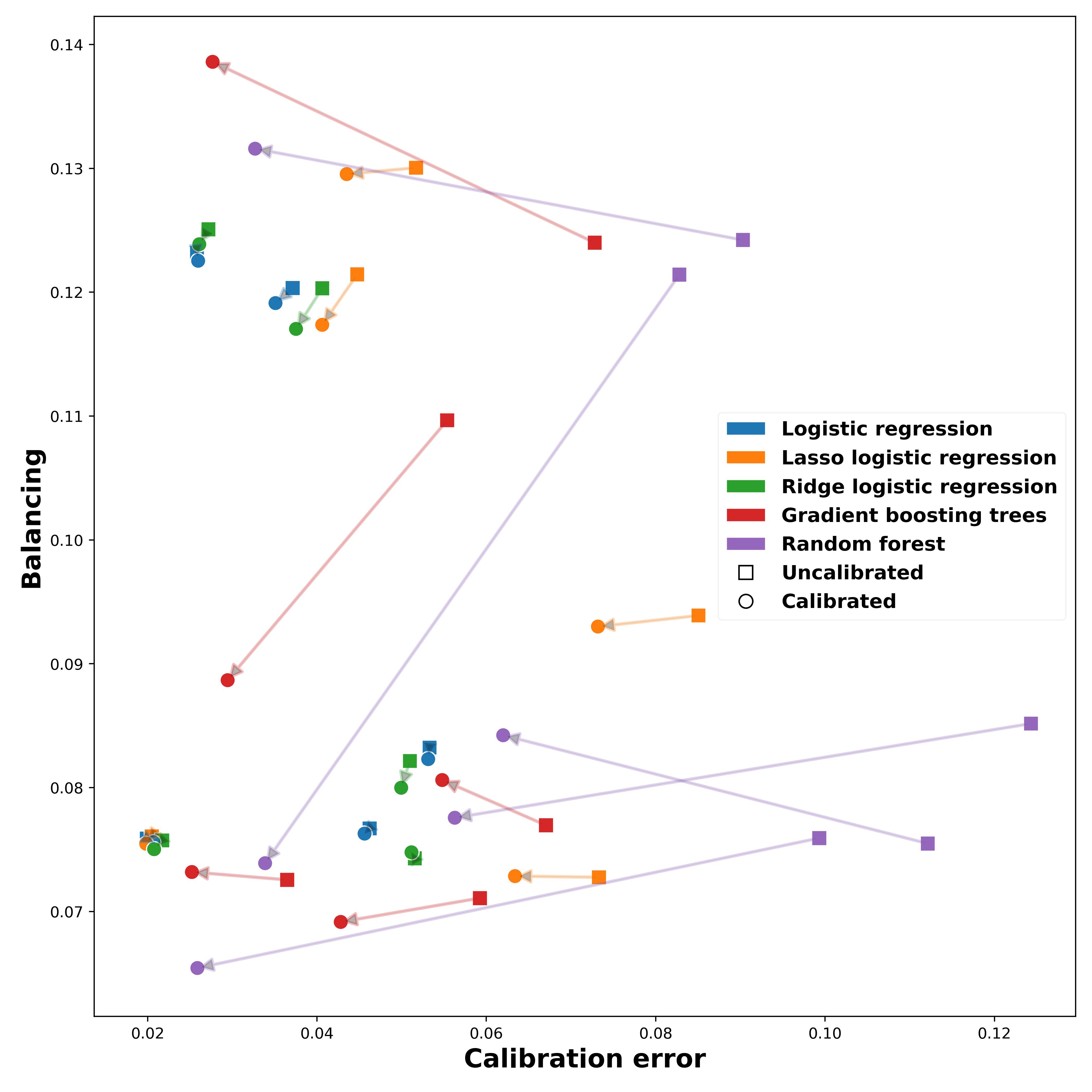}
        \caption{}
        \label{fig:acic_cal_bal}
    \end{subfigure}
    \caption{Post calibration improves effect estimation, independent of its effect on balancing.
    Different statistical estimators (colors) were used to estimate propensity scores 
    on a dataset from the 2016 Atlantic Causal Inference Conference Data Challenge,
    before (rectangle marker) and after (circle) post-calibration with a connecting arrow.
    \textbf{a)} Calibration error (mean difference if Integrated Calibration Index) is plotted against the absolute difference in between the true and estimate average treatment effect.
    Post-calibrated scores achieved lower effect estimation error, 
    while not necessarily improving balancing (measured as maximum value of absolute standardized mean difference of covariates in the right panel (b)). 
    }
    \label{fig:acic_cal_scatter}
\end{figure}

Figure \ref{fig:acic_cal_scatter} (a) shows that post-calibration consistently reduces the error between the true and estimated average treatment effect.
In addition, similarly to before, calibration doesn't seem to have a consistent effect on balancing.

\begin{figure}
    \centering
    \includegraphics[width=\textwidth]{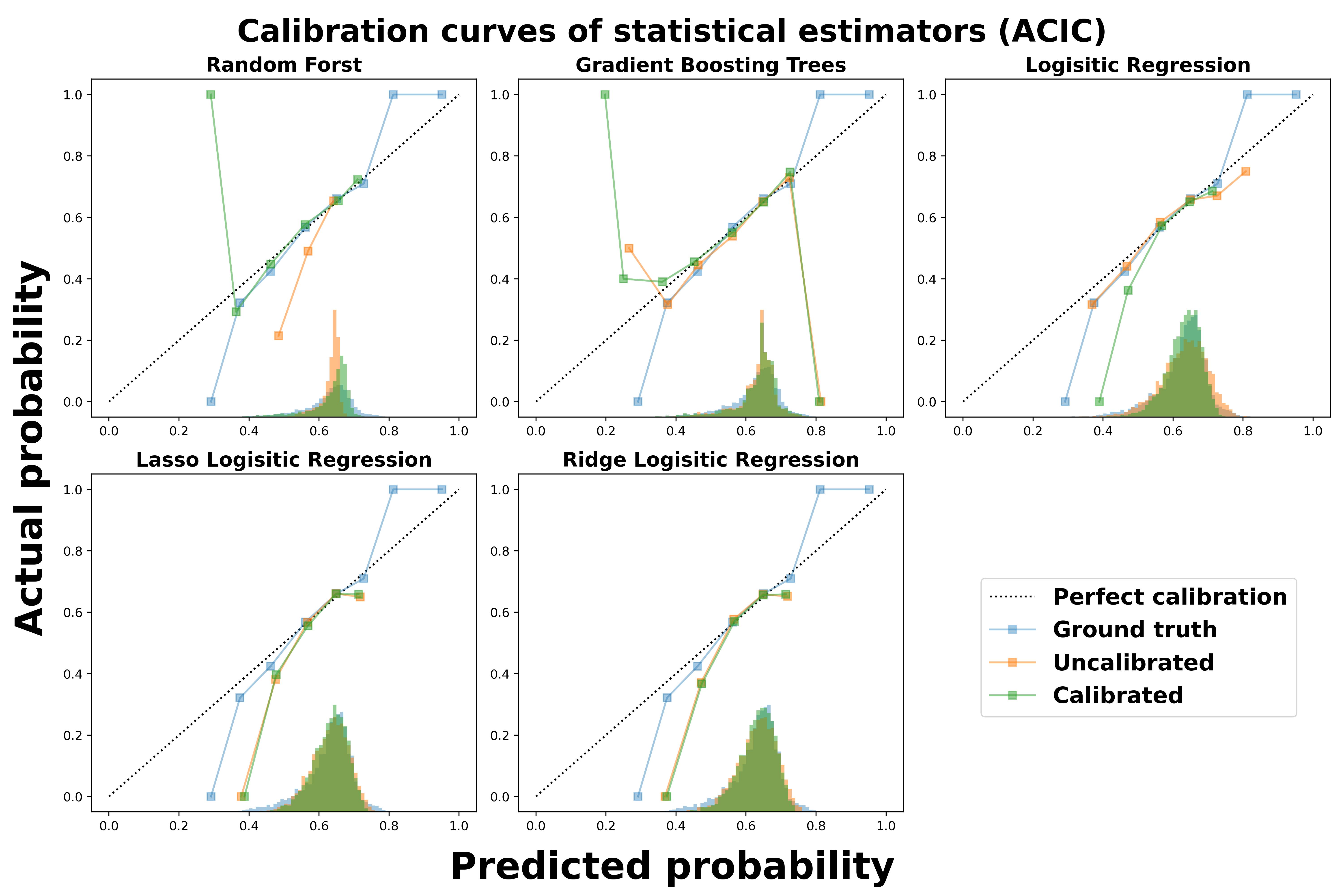}
    \caption{Calibration curves of different model on a single ACIC dataset.
    The true simulated propensity scores (blue), not optimally on the diagonal since the tails of the distribution are thin and variance is high.
    Original scores from the estimators (orange), are at least as farther from the diagonal, since they are estimated by a binary instantiation of the true propensities.
    Post-calibrated scores (green) are slightly closer to the diagonal relative to the original scores, as expected. 
    }
    \label{fig:acic_cal_curve}
\end{figure}

The calibration curves from one specific data set, shown in Figure \ref{fig:acic_cal_curve}, tell a similar, yet less decisive, story to the model-free propensity simulation one in Figure \ref{fig:sim_cal_curve} (and the model-based results applied on the simpler data in Figure \ref{fig:sim_model_cal_curves}).
Out of the box, the models (orange lines in each panel) are not very calibrated, diverging from the x=y diagonal (dashed line).
However, post-calibrating the propensity scores does improve calibration (green line) moving the curve closer to the diagonal. 

\section{Discussion}
We performed a simulation study to assess the effect of post-calibrating propensity scores on downstream estimation of causal effects.
We've shown that miscalibrating of propensity score models results in poor effect estimation and that post-calibrating models improves effect estimation.

We rely on simulated data since we needed access to two variables that are usually unobserved. 
The first is the individual's true propensity to be treated.  
The second is the counterfactual outcomes, in order to calculate the true average treatment effect and compare it to the estimated effect.

There are various ways to use propensity scores for adjustment.
Three main ones are through matching, stratification and weighting.
In this study we chose inverse probability weighting (IPW), since it has fewer degrees of freedom than the others.
Stratification requires an additional binning parameter and matching requires transforming the data set, discarding samples, and usually estimating average treatment effect on the treated.
Weighting, on the other hand, provides a smoother and more continuous transformation of the propensities, and therefore is subjected to less modeling artifacts.


Post-calibration (re-calibration) is the process of taking the scores provided by an estimator and transforming them so they behave more like probabilities.
Overall, post-calibrating propensity scores had consistently improved effect estimation
by reducing bias.
Figure \ref{fig:sim_cal_deform} shows this in a theoretical scenario where the true propensity scores were gradually deformed and then re-calibrated. 
We do that to start with a tightly-controlled dose-response-like analysis.
We show that worsening the calibration results in worse treatment effect estimation,
and that post-calibration resulted in lower estimation error.
Figure \ref{fig:acic_cal_scatter} gets to the same conclusion in a more realistic scenario where actual statistical models are used. 
Since not all models are properly calibrated out-of-the-box, post-calibrating them improves downstream effect estimation.

Furthermore, this improvement in estimation due to calibration is not mediated by improved balancing.
Namely, Figures \ref{fig:sim_cal_bal} and \ref{fig:acic_cal_bal} show that calibration does not consistently improve balancing, while Figures \ref{fig:sim_cal_ate} and \ref{fig:acic_cal_ate} show it consistently improves effect estimation.
Therefore suggesting it is the better calibration that has improved the effect estimation, regardless of whether or not balancing was also improved in the process. 

Not all models gain the same benefit from post-calibration.
Logistic regression and its regularized variants tend to improve less than their tree-based counterparts.
This phenomena is in line with regression-based models being more calibrated to begin with, by optimizing log-loss or a biased log-loss.
However, tree-based models seem to benefit substantially,
with some instances reaching smaller error after calibration than the logistic-based models.
This seems to break the trade-off between model expressiveness and model calibration, allowing both to exist simultaneously. 
Hence, supporting the claim that if more expressive models are required for modeling the exposure, they are probable to benefit from post-calibration. 

Since "the fundamental problem of causal inference" forces us to use simulations for this study,
how much our conclusions generalize is also dependent on how much of the properties of these simulated data also exist in real data.
We therefore showed results of both a simple case and a complex case.
In addition, estimators are still sensitive to underfit and overfit.
Underfitted propensity models fail to capture the signal of the treatment assignment mechanism, resulting in their propensity scores being less informative to begin with, and therefore might not benefit from post-calibration like properly specified models.
Conversely, overfitted propensity models will predict treatment assignment so well that the distribution of their scores will differ across treatment groups, 
making overfitted models indistinguishable from positivity violations.
Hence preventing us from converting our statistical estimations into causal claims.
However, these issues are not unique to propensity model, but apply to general prediction models as well, 
and can be partly automated through hyperparameter search and cross validation.

Limitations notwithstanding, post-calibrating estimators allows to reconcile theory - guarantees relying on the true conditional probability of exposure - and practice - as in applied numerical modeling in statistical software.
Post-calibration is a simple postprocessing procedure, available in common statistical software, that can be done on any statistical estimator.
It is usually not computation-intensive,
allowing us to utilize more complex models at a relatively small additional cost.
Therefore, we conclude that post-calibrating propensity score models can be beneficial for effect estimation.

\bibliographystyle{unsrtnat}  
\bibliography{references}

\newpage

\newcommand{\beginsupplement}{%
        \setcounter{table}{0}
        \renewcommand{\thetable}{S\arabic{table}}%
        \setcounter{figure}{0}
        \renewcommand{\thefigure}{S\arabic{figure}}%
     }

\appendix
\beginsupplement
\section*{Supplementary Materials} \label{sec:supp_mat}

\subsection*{Estimation-free propensity scores}
In Figure \ref{fig:sim_cal_deform}, we show the results for ten different random datasets per deformation scale.
To obtain more rigorous statistics, we repeat that process a thousand times per deformation scale, rather than ten.
For each such instance we calculate the slope between the pre- and post-calibration point on the calibration-error effect estimation error plane.
The summary of a 1000 such slopes per deformation scale is presented in Table \ref{table:sim_deform}.
In which we can see two things.
First, the larger the deformation - on each side of the deformation scale (below and above 1) - the larger the magnitude of the slope is.
Second, the negative sign suggests post-calibration constantly improves reduces calibration error and effect estimation error.
For "deformation" scale 1.0, which is no deformation at all, we would expect a slope of zero. 
However we do see a small median slope of minus one, but with a large interquantile range. 
This hints that there isn't a truly significant slope there, but rather it is quantifying the noise.

\begin{table}[h]
\centering
\begin{tabular}{ |c|r|r| r| }
\hline
Deformation scale & 1st quartile & Median & 3rd quartile \\
\hline
2.0               & -54.10       & -39.40 & -25.06       \\
1.75              & -52.43       & -36.47 & -21.51       \\
1.5               & -96.84       & -56.91 & -18.41       \\
1.0               & -20.07       & -1.03  & 23.10        \\
0.75              & -9.31        & -7.41  & -0.74        \\
0.5               & -10.08       & -8.91  & -5.89        \\
0.25              & -10.83       & -9.95  & -8.04       \\
\hline
\end{tabular}
\caption{Slopes between pre- and post-calibration points on the calibration-error over effect-estimation error plane. Statistics are taken for a 1000 datasets per deformation scale.
First, the larger the deformation the larger the magnitude of the slope is.
Second, the negative sign suggests post-calibration constantly improves reduces calibration error and effect estimation error.}
\label{table:sim_deform}
\end{table}

\subsection*{Model-estimated propensity scores on simple simulation} \label{sec:syn_models}
\begin{figure}[!h]
    \centering
    \includegraphics[width=\textwidth]{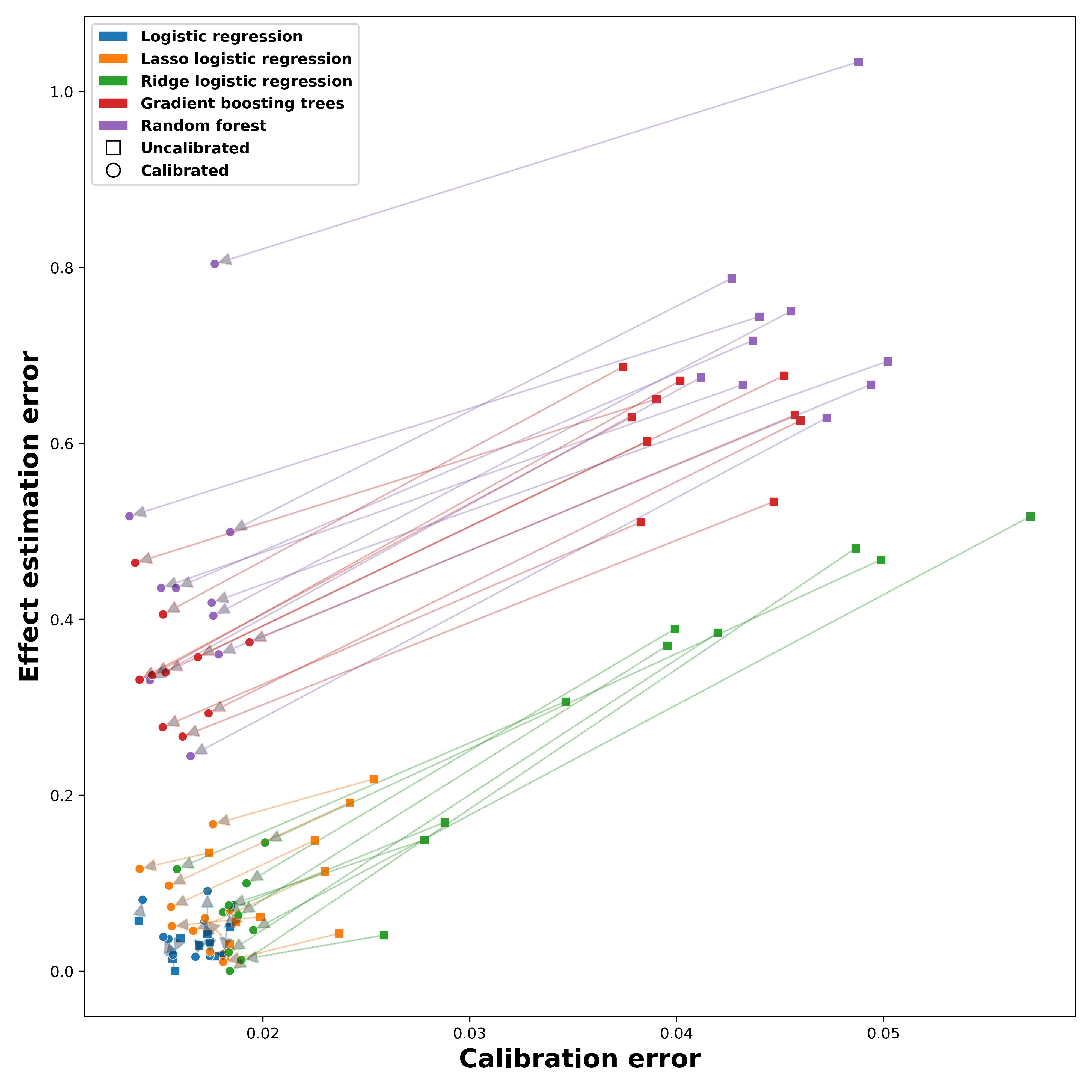}
    \caption{Calibration improves effect estimation in a simple simulation scenario.}
    \label{fig:sim_model_cal_ate}
\end{figure}

We also applied statistical estimators the data generating process described in \nameref{sec:methods} section, setting $\gamma = 1$. 
Results in Figure \ref{fig:sim_model_cal_ate} show that the improvement of calibration is consistent,
similar to what presented on the 2016 Atlantic Causal Inference Conference data (Figure \ref{fig:acic_cal_scatter}). 
However, we do see that the logistic regression model has a small and inconsistent benefit.
That might be due to the model, already being perfectly specified to the data and well-calibrated, doesn't benefit from additional logistic regression-based calibration.
We also see the tree-based models perform worse than the logistic regression-based models.
This is probably due to the data being generated via regression, and trees being less adequate to fit linear trends (given finite complexity). 

\begin{figure}[t]
    \centering
    \includegraphics[width=\textwidth]{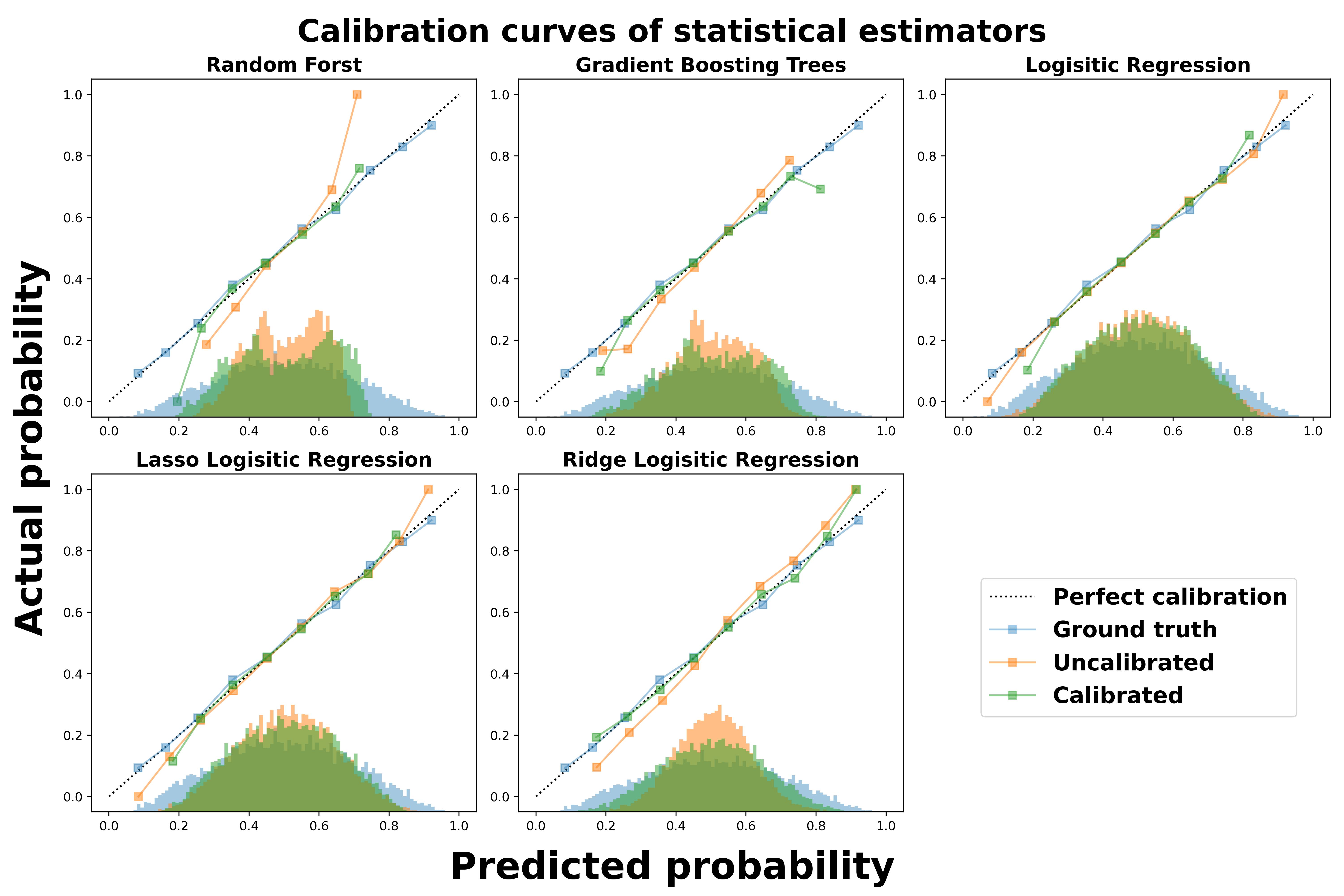}
    \caption{Post calibration improves calibration curves when applying different statistical estimators on a simple simulation data.}
    \label{fig:sim_model_cal_curves}
\end{figure}

To further see the effect of post-calibration we choose an arbitrary simulation instance and plots its corresponding calibration curves. 
in Figure \ref{fig:sim_model_cal_curves} we see the post-calibrated curves (green) are closer to the optimal x=y diagonal than the original curves.
We also see the post-calibrated score distributions are more spread to match the ground truth distribution.
The above conclusion is limited to the logistic regression model, since the simple simulation uses a logit function of a linear combination, logistic regression is perfectly well-specified,
and therefore, shows little change following the post-calibration procedures.

\end{document}